\documentclass[useAMS,usegraphicx]{mn2e}
\usepackage{times}
\newcommand{\beq}{\begin{equation}}
\newcommand{\eeq}{\end{equation}}

\newcommand{\chir}{\chi_{\rm r}}

\newcommand{\cs}{c_{\rm s}}

\newcommand{\Msun}{~M_{\odot}}
\newcommand{\td}{t_{\rm d}}

\newcommand{\Pe}{P_{\rm ext}}
\newcommand{\Mc}{M_{\rm c}}
\newcommand{\Ms}{M_{\rm s}}
\newcommand{\Mu}{M_0}

\begin{document}

\title[Power-Law Tail]{On the Power-Law Tail in the Mass Function of Protostellar Condensations and Stars}
\author[Shantanu Basu and C. E. Jones]{Shantanu Basu$$\thanks{E-mail: basu@astro.uwo.ca (SB)} and C. E. Jones$$\\ 
Department of Physics and Astronomy, University of
Western Ontario, London, Ontario, Canada N6A 3K7}


\maketitle

\label{firstpage}

\begin{abstract}
We explore the idea that the power-law tail in the mass function of 
protostellar condensations and stars arises from the 
accretion of ambient cloud material on to a condensation,
coupled with a nonuniform (exponential) distribution of 
accretion lifetimes.  This model 
allows for the generation of power-law distributions in all star-forming
regions, even if condensations start with a lognormal
mass distribution, as may be expected from the central limit theorem,
and supported by some recent numerical simulations of turbulent 
molecular clouds.
For a condensation mass $m$ with growth rate $dm/dt \propto m$, an analytic
three-parameter probability density function is derived; it resembles a 
lognormal at low mass and has a pure power-law high-mass tail. An approximate
power-law tail is also expected for other growth laws, and we 
calculate the distribution for the plausible case $dm/dt \propto m^{2/3}$.
Furthermore, any single time snapshot of the masses of 
condensations that are still accreting (and are of varying ages) also 
yields a distribution with a power-law tail similar to that of the IMF.

\end{abstract}


\section{Introduction}

\subsection{IMF background}

Despite decades of detailed modeling of
gravitational collapse and fragmentation, it is still not
clear how certain physical conditions determine the mass of an individual 
star; a larger question is the origin of the {\it distribution} of 
stellar masses at birth, the stellar initial mass function (IMF). 
Additionally,
the observed mass spectrum of protostellar condensations 
in molecular clouds appears to follow a power-law decrease,
similar to that of stars, in the mass range above $\sim 0.5 \Msun$
(Motte, Andr\'e, \& Neri 1998; Testi \& Sargent 1998; Johnstone et al.
2000).
Hence, the fundamental shape of the IMF may be determined at 
the molecular fragment scale, even though the transformation of
a condensation to star(s) is not a simple process.

Starting with the expression for the Jeans (1928) mass, and working up to
more sophisticated scenarios, there is a general tendency for multiplicative
dependence on physical parameters (see, e.g., Adams \& Fatuzzo 1996), leading
to the claim that the central limit theorem should ensure that
the IMF has a lognormal form. The formation
of star-forming condensations in a molecular cloud is surely affected
by multiple effects of a turbulent flow field, and statistical
variations of ambient magnetic field, density, pressure, and temperature. 
Numerical models of turbulent molecular clouds typically yield a lognormal
distribution of dense clump masses (e.g., Padoan, Nordlund, \& Jones 1997;
Ostriker, Stone, \& Gammie 2001; Klessen 2001).  
Although much additional physics remains to be included in these models, 
it may be reasonable to believe that the stochastic feature of a lognormal
distribution will continue to hold.
Hence, one can ask a simple set of questions about star formation:
Is the IMF lognormal? If not, and an approximate power-law tail is 
inferred, then how is it generated?

While Miller \& Scalo (1979) did obtain a lognormal fit to the IMF, 
subsequent detailed studies find a region
of power-law decrease in the intermediate and high mass ($M$) range that is 
similar to the original result of
Salpeter (1955), i.e., $dN/d \ln M \propto M^{-1.35}$, 
where $dN$ is the differential number of stars
in the logarithmic mass interval $d \ln M$ (see e.g., reviews by
Scalo 1998; Meyer et al. 2000; Elmegreen 2001).
Scalo (1998) emphasizes a varying power-law index in the 
intermediate and high mass range, with significant 
region to region variations reflecting intrinsic variability 
and/or observational uncertainties.

\subsection{Lognormal or power-law distribution?}

The generation processes of lognormal and power-law 
distributions are often intimately connected. This phenomenon has been 
studied extensively in social science, biology, and computer science,
e.g., in the study of power-law distributions of income (the Pareto Law), 
city sizes (Zipf's Law), and Internet file sizes. Useful models and 
discussion can be found in Reed (2002, 2003), Reed \& Hughes (2002),
Mitzenmacher (2003), and references within.
It turns out that many seemingly small effects in the
generative process of a lognormal distribution can lead to
a power-law tail instead.
Such an explanation seems particularly relevant for the IMF, since
the power-law tail is typically measured to span even less than 
two decades of mass, hardly qualifying as evidence for a truly
scale-free process.


In this Letter, we propose a minimum-hypothesis explanation:
even if the central limit theorem is relevant in determining the starting 
masses of condensations, the temporal effect of their 
subsequent interaction with the environment skews the mass 
distribution into a power-law form. 
This paper is based on the statistical model summarized by Reed \& 
Hughes (2002), and builds upon some elements of the model of Myers (2000).
The basic idea is that 
the growth of masses by accretion coupled with a {\it distribution of times}
for the growth yields a power-law tail in the mass function. 
It is a statistical approach,
not dependent on the detailed values of physical variables in 
different environments. While the mass of any individual star is
dependent on the particular initial values of various physical parameters,
we are concerned here only with the {\it shape} of the mass distribution
function; specifically the intermediate and high mass tail.



\section{Lognormal Fits and Power-Law Tail in the IMF: A Case Study}

\begin{figure}
\includegraphics[width=84mm]{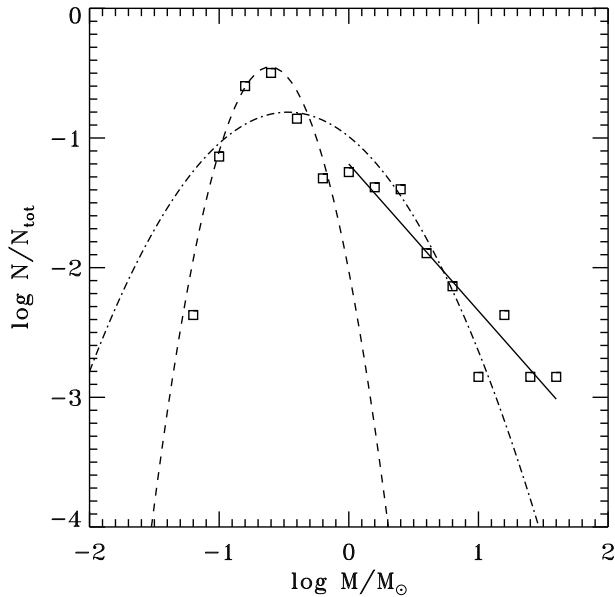}
\caption{Estimated stellar masses in the ONC and
best-fitting analytic curves. The squares represent the masses estimated
by H97, binned in intervals $\Delta \log (M/M_{\odot}) = 0.2$.
The dashed line is the best-fitting lognormal distribution (eq. [\ref{lognorm}],
with $\mu=-1.40, \sigma=0.52$) for $M < 1 \Msun$, the dash-dotted line
is the best-fitting lognormal distribution ($\mu = -1.08, \sigma = 1.16$)
for all masses. The solid line is the least squares best-fitting straight line  
for the region $M \geq 1 \Msun$; it has slope $\Gamma = d\log N/d\log M 
= -1.13 \pm 0.15$.
}
\label{fig1}
\end{figure}

In order to establish the minimum requirements for an IMF model,
we review the data from the Orion Nebula Cluster (ONC)
obtained by Hillenbrand (1997, hereafter H97). 
The ONC is a unique laboratory for the study of the IMF due to  
its relatively nearby location and direction away from the Galactic plane, 
allowing a deep sampling of the young stellar population in the vicinity 
of the massive star $\theta^1 C$. The large range of surface density
from centre to observed edge of the cluster, and the
large number of observed stars allows for useful statistics on the role
of the environment (specifically the density of stars) in the generation
of the IMF. Furthermore, since most stars have estimated ages less than a
few Myr (H97), they are likely observed near their birth sites.
The IMF of H97 also shows a decline at lower masses, which
we utilize to assess the viability of lognormal fits. 

Figure 1 shows a histogram of 696 stellar masses obtained by H97;
these are the subset of observed stars which are reported to have a 70\% or 
higher chance of membership in the ONC.
We do not try to account for uncertainties in the mass determinations here,
and simply seek a best-fitting lognormal probability density function
\beq
\label{lognorm}
f(m) = \frac{1}{\sqrt{2 \pi} \sigma m} \exp \left[ - \frac{(\ln m - \mu)^2}
{2 \sigma^2} \right] 
\eeq
for the normalized masses $m=M/M_{\odot}$. For comparison with a sample
of $N_{\rm tot}$ objects,
the normalized number in each bin, $dN/N_{\rm tot}$, equals
$f(m) dm = m f(m) d \ln m$, where $d \ln m$ is the width of a logarithmic
mass bin.
We conduct a parameter search of values of the mean $\mu$ and
variance $\sigma^2$ of $\ln m$. Minimization of the $\chi^2$ deviation of
the observed values from the analytic curves 
yields best-fitting parameters $\mu = -1.08$ and
$\sigma = 1.16$. The dash-dotted line in Figure 1 is a lognormal
distribution with these parameters. Clearly, 
the ability to fit the tail of the distribution is inconsistent
with fitting the peak at the low mass end. The reduced $\chi^2$ for this
fit is $\chir^2 = \chi^2/\nu = 6.64$ for $\nu = 12$ 
degrees of freedom, yielding a probability 
$4.75 \times 10^{-12}$ that the data is drawn from the lognormal distribution. 
Alternatively, we
may seek the best-fitting lognormal to a subset of the data; 
fitting only stars in bins centred at 
$M < 1 \Msun$ yields best parameters
$\mu = -1.40$ and $\sigma = 0.52$. Figure 1 also shows this fit 
(dashed line), which has $\chir^2 = 1.86$ and $\nu=3$,
yielding a probability $0.134$ that the data is drawn from the distribution.
A conclusion is that the IMF can be approximately fit by a lognormal 
at low mass, but has a power-law intermediate and high mass tail. 
The histogram yields a least squares best-fitting straight line
of slope $\Gamma = d\log N/d\log M = -1.13$ for $M \geq 1 \Msun$.


Due to the large variation in stellar surface density in the ONC,
we also look
for systematic location-dependent effects on the IMF by dividing the
stars into three radial zones, each containing about 232 stars,
enough for a meaningful statistical sample.
Table 1 lists the values of $\chir^2$ from comparing the 
histogram of masses $M < 1 \Msun$ in each zone with the best-fitting lognormal
for all zones: $\mu = -1.40$, $\sigma = 0.52$. The implied 
probabilities range from 0.287 to 0.616.
For $M \geq 1 \Msun$, the best-fitting slopes to
the power-law tail are shallower than $-1$ in radial zones 1 and 2,
although there are significant uncertainties in the 
fits.  Interestingly, if we exclude the (merely five)
stars in bins centred at $M > 10 \Msun$, all slopes become somewhat
steeper than $-1$.
Four of the five excluded stars are in zone 1 (these are 
the Trapezium stars) and one is in zone 2.

\begin{table*}
 \centering
  \begin{minipage}{140mm}
 \caption{Summary of best-fitting parameters.} 
\begin{tabular}{@{}lccccr@{}}
  \hline
& & 
\multicolumn{2}{c}{$\rm M\,<\,1\, M_{\odot}$} &
{$\rm M\,\geq \,1 \, M_{\odot}$ }&{ $1 \, \rm M_{\odot} \leq M \leq 10 \,\rm M_{\odot}$}\\
& & \multicolumn{2}{c}{\hrulefill} &
{\hrulefill} & {\hrulefill}\\
Zone & Surface Density (${\rm pc}^{-2}$) &
lognormal $\chir^2$ & probability & best slope & 
best slope \\
\hline

all ($0 \leq r < 17.70$ pc) & 0.71 & 1.86 & 0.134  &$-1.13\pm 0.15$ & $-1.53\pm 0.27$ \\
1 ($0 \leq r < 4.32$ pc)    & 3.94 & 0.598 & 0.616  & $-0.85\pm 0.21$  & $-1.36\pm 0.39$\\
2 ($4.32 \leq r < 8.65$ pc) & 1.32  & 1.25  & 0.287  &$-0.89\pm0.16$ & $-1.12\pm 0.22$\\
3 ($8.65 \leq r < 17.70$ pc) & 0.31 & 1.17 & 0.318  &$-1.10\pm0.45$ &$-1.10\pm0.45$\\
\hline
\end{tabular}
\end{minipage}
\end{table*}

The ONC data implies a generation of essentially the same
power-law in {\it all} regions, despite large differences in stellar density
(see Table 1). 
The five most massive stars, four of which are very near the core of the ONC, 
provide the only significant evidence for a region-dependent IMF.
Their presence and location may be due to events occurring after
star formation, since the local relaxation time decreases
toward the cluster centre due to increasing density. However, the 
simulations of dynamical mass segregation by Bonnell \& Davies (1998) 
imply that they could not have formed very far from the cluster centre. 
In any case, we bypass the nagging questions about the origin of these five 
stars and focus instead on an explanation for the generation of a universal 
power-law tail in the mass function of all other stars.

\section{A Statistical Model of Growth by Accretion}

\subsection{Basic model}

\begin{figure}
\includegraphics[width=84mm]{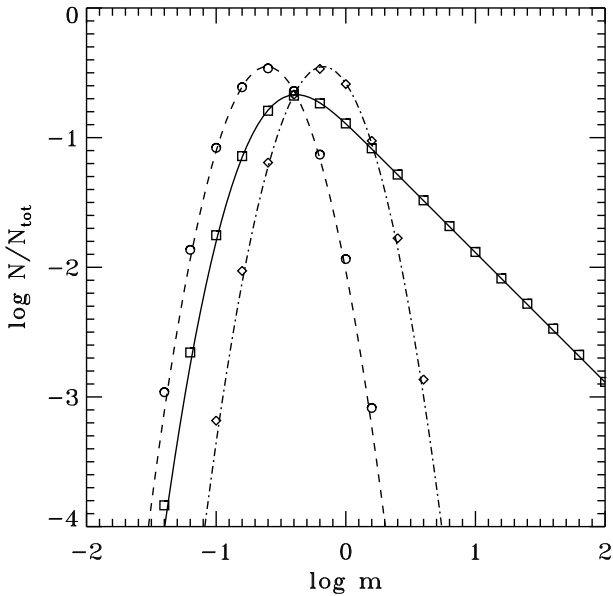}
\caption{Evolution of a distribution of normalized masses $m$ 
undergoing accretion growth $dm/dt \propto m$ (eq. [\ref{multmdot}]).
The dashed line is
an example initial lognormal distribution with $\mu_0=-1.40,\sigma_0=0.52$,
and the circles represent a histogram (binned with $\Delta \log m = 0.2$)
of a random sample of $10^6$ objects drawn from the distribution.
The dash-dotted line is the analytic expression for the distribution
(eq. [\ref{lognorm}] with $\mu = \mu_0+1$) if the masses grow 
for a fixed time $t = \gamma^{-1}$. 
The solid line is the analytic density function (eq. [\ref{newdistfcn}]
with $\mu_0=-1.40,\sigma_0=0.52, \alpha = 1$) 
if accretion lifetimes have density function
$f(t) = \delta e^{-\delta t}$, and $\delta = \gamma$.
The diamonds and squares represent the histogram of the 
masses in the original distribution if they grow
for a fixed time or with an exponential distribution of times, respectively; 
they necessarily overlap the analytically predicted curves.
}

\label{fig2}
\end{figure}

According to the central limit theorem of statistics, if 
the mass of a protostellar condensation $\Mc = f_1 \times f_2 \times ... 
\times f_N$, then the distribution
of $\Mc$ tends to a lognormal regardless of the distributions of the
individual physical parameters $f_i$ ($i=1,...,N$), if $N$ is large. 
Depending on the specific distributions of the $f_i$, a convergence to a 
lognormal may even occur for moderate $N$ (Ioka \& Nakamura 2002).
However, the stellar IMF results from further evolution beyond the
condensation phase, so that the mass of a star may be characterized 
as $\Ms = \Mc \times f_{\rm frag} \times f_{\rm out}$, where 
$f_{\rm frag}$ is the
fraction of the condensation mass that goes into a star after any 
possible fragmentation
process, and $f_{\rm out}$ is the fraction of mass that remains after 
mass loss due to a protostellar outflow. The multiplicative nature
of these processes seems consistent with theoretical models (Shu et al.
1999; Meyer et al. 2000),
and imply that the distribution of $\Ms$ should be lognormal if
that of $\Mc$ is lognormal as well.

Here we explore the idea that the power-law tail is 
due to the interaction of a condensation with its
environment in the time before it collapses down to stellar 
dimensions.
The environment of a condensation is certainly a large reservoir of
mass, and there may be dynamical influences due to the passage
of ambient pressure fluctuations, including shocks.
If the accretion is a continuous time multiplicative process,
\beq
\label{multmdot}
\frac{dm}{dt} = \gamma m \;  \Rightarrow  \; \: m(t) = m_0 \exp(\gamma \, t),
\eeq
where $\gamma$ is a growth rate and $m_0$ is an initial mass. If
the values of $m_0$ are drawn from a lognormal distribution with 
mean $\mu_0$ and variance $\sigma_0^2$, the distribution of masses
$f(m)$ at any later time $t$ will still be described by 
equation (\ref{lognorm}), but with
\beq
\mu = \mu_0 + \gamma \, t, \: \: \sigma = \sigma_0.
\eeq

However, all condensations may not accrete for
the same time period, and a distribution of accretion times $t$
will in fact skew the final distribution away from a lognormal.
A simple example of a distribution of times is if the 
probability of stopping accretion is constant in time. A 
constant ``death'' rate $\delta$ for accretion leads to an 
exponential distribution, with probability density function
$f(t) = \delta e^{-\delta t}$.
In this case, the probability density function for masses is 
\beq
\label{timeconv}
f(m) = \int_{0}^{\infty} \frac{\delta e^{-\delta \, t}}{\sqrt{2 \pi} \sigma_0 m}  \exp\left[-\frac{(\ln m - \mu_0 - \gamma \, t)^2}{2 \sigma_0^2}  \right]  dt.
\eeq
Using the integral identity
\[
\int_{0}^{\infty} \exp \left[ -(ax^2 + bx + c) \right] dx  
\]
\beq
 =  \frac{1}{2} \sqrt{\frac{\pi}{a}} \exp \left[ (b^2 - 4ac)/4a \right] {\rm erf
c} \left( \frac{b}{2 \sqrt{a}} \right),
\eeq
in which
\beq
{\rm erfc}(x) = \frac{2}{\sqrt{\pi}} \int_{x}^{\infty} \exp(-u^2) \, du
\eeq
is the complementary error function, we find 
\begin{eqnarray}
\label{newdistfcn}
f(m)  & =  & \frac{\alpha}{2} \exp \left[ \alpha \mu_0 + \alpha^2\sigma_0^2/2 \right] \: m^{-1 - \alpha}  \nonumber \\
& & \times \: {\rm erfc} \left[ \frac{1}{\sqrt{2}} \left( \alpha \sigma_0 - \frac{\ln m -\mu_0}{\sigma_0} \right) \right], 
\end{eqnarray}
where $\alpha = \delta/\gamma$ is the dimensionless ratio of ``death'' rate to 
``growth'' rate of condensations. Equation (\ref{newdistfcn}) represents
a new three-parameter ($\mu_0,\sigma_0,\alpha$) probability density 
function which
tends to the form $m^{-1-\alpha}$ for large masses, but is modulated by
the complementary error function (which goes to 0 for very low $m$ and 
to 2 for large $m$) so that it tends toward zero at very low masses.
It is therefore similar to a lognormal distribution with a power-law tail.
(Note that it reduces to a pure power-law distribution if all
stars start with the same mass $m_{\rm i}$, i.e., $m_0$ is described by a 
delta function, so that $\mu_0 = \ln m_{\rm i}$ and $\sigma_0=0$ in 
eq. [\ref{newdistfcn}])
We believe this function can be used to fruitfully model the IMF. 

Physically, we expect $\alpha \approx 1$ since both $\delta$ and $\gamma$
are rates controlled by the external medium and are expected to be 
approximately the 
inverse external dynamical time $\td^{-1}  = (G \rho_{\rm ext})^{1/2}$, 
in which $\rho_{\rm ext}$ is the mean density outside the condensation.
Myers (2000) has derived an example of exponential growth with $\gamma
\approx \td^{-1}$, assuming geometric accretion (see eq. [\ref{geomacc}])
and that every condensation increases its internal turbulence
continuously (as the mass increases) exactly as needed to remain in a
critical Bonnor-Ebert state.

Figure 2 shows a sample initial lognormal distribution 
($\mu_0 = -1.40, \sigma_0 = 0.52$), and a distribution
at a later time under the assumptions (1) multiplicative growth
(eq. [\ref{multmdot}]) in which all condensations accrete for a fixed time
$t = \gamma^{-1}$, and (2) multiplicative growth with an exponential
distribution of times $f(t) = \delta e^{-\delta t}$. Under assumption (1), 
the new distribution is again a lognormal with $\mu = \mu_0 + 1$, but 
in case (2) it is given by equation (\ref{newdistfcn}) and has a 
power-law tail with index $\alpha=1$ if $\delta = \gamma$.

Recently, Reed (2002, 2003) has derived a probability density function that
is even more general than equation (\ref{newdistfcn}). It assumes the 
growth law known as Geometric Brownian Motion:
$dm/m = \gamma \, dt + \sigma_1\, dw$, where $dw$ is a 
``white-noise'' random number drawn uniformly from the range $[0,1]$. 
In this case, an initially lognormal
distribution characterized by $\mu_0$ and $\sigma_0$ also remains lognormal 
after a fixed time $t$, but with
variance $\sigma^2 = \sigma_0^2 + \sigma_1^2\,t$, and
mean $\mu = \mu_0 + (\gamma - \sigma_1^2/2)\, t$.
For an exponential distribution of 
lifetimes, Reed derives an analytic 
four-parameter density function, more complicated than equation 
(\ref{newdistfcn}), but which also has the asymptotic dependence 
$f(m) \propto m^{-1-\alpha}$ for large $m$.
We refer the reader to the above-mentioned papers for details.


While an exponential distribution of accretion termination times guarantees
a power-law tail in the mass distribution, it is also true that all 
condensations need not {\it start} accreting at the same time. 
If the ``birth''
of condensations is distributed uniformly in time throughout the life of
a cloud, and termination times obey an exponential distribution, then the
number of ``living'' (i.e., still accreting) condensations at a fixed time $t$
of observation
will also obey an exponential distribution of times $\tau$ since birth, i.e.,
$f(\tau) = \delta e^{-\delta \tau}$. Furthermore, if 
the ``birth'' of condensations is also increasing exponentially in time
with a growth rate $\beta$, then the distribution of 
times $\tau$ since birth 
will be $f(\tau) = (\beta + \delta) e^{-(\beta+\delta)\tau}$.
Hence, observations of accreting condensations made at a single 
snapshot in time, 
e.g., Motte et al. (1998), should yield a power-law index 
$\alpha' = \alpha + \beta/\gamma$, so that the slope is equal to 
that of the IMF if $\beta=0$ (or $\beta \ll \gamma$), but steeper otherwise.

\subsection{A variation on the basic model}

\begin{figure}
\includegraphics[width=84mm]{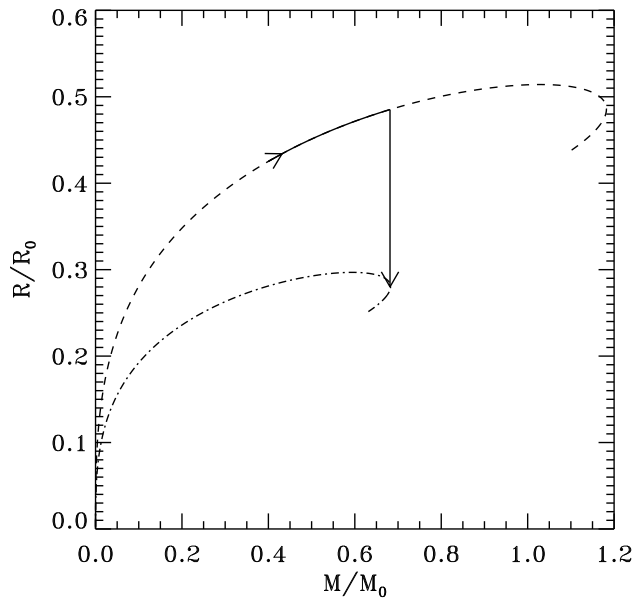}
\caption{Radius ($R/R_0$) versus mass ($M/M_0$) relations for a sequence 
of Bonnor-Ebert equilibrium
states for constant external pressure $\Pe$ (dashed line). Here,
$R_0 = \cs^2/(G^{1/2}\Pe^{1/2})$ and 
$M_0 = \cs^4/(G^{3/2} \Pe^{1/2})$. The plotted 
sequence extends somewhat beyond the
critical mass $M_{\rm crit} = 1.18 \, M_0$.
The dash-dotted line is the corresponding sequence for an external
pressure $3\, \Pe$. The solid line with arrows represents a possible 
evolutionary sequence of a condensation during mass accretion, terminating
with collapse as the external pressure suddenly rises to a value incompatible
with an equilibrium state.
}
\label{fig3}
\end{figure}

\begin{figure}
\includegraphics[width=84mm]{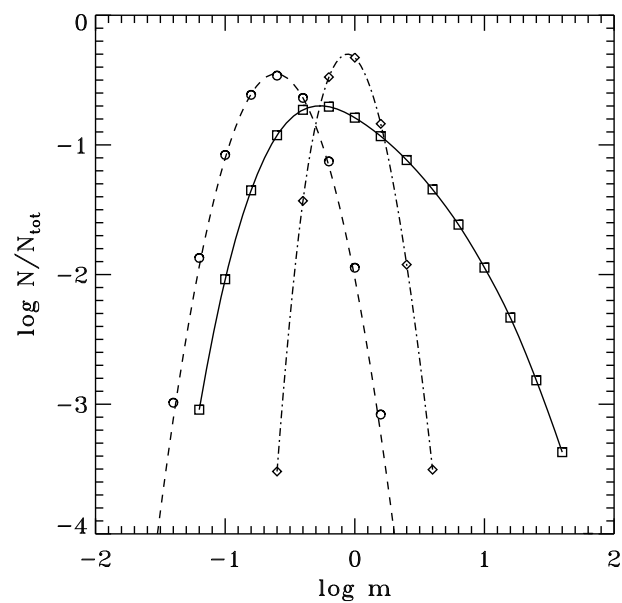}
\caption{Evolution of a distribution of normalized masses $m$ 
undergoing accretion growth $dm/dt \propto m^{2/3}$ (eq.[\ref{bemdotandsoln}]).
The dashed line is 
an example initial lognormal distribution with $\mu_0=-1.40,\sigma_0=0.52$.
A sample of $10^6$ masses drawn from this distribution are binned
in increments $\Delta \log m = 0.2$ and represented by circles. 
The diamonds (connected by a dash-dotted line)
are a histogram of masses after they have all accreted for a 
time $t = \gamma_1^{-1}$, and the squares (connected by a solid line) 
are a histogram of masses if 
the accretion lifetimes have density function
$f(t) = \delta e^{-\delta t}$, in which $\delta = \gamma_1$.
}
\label{fig4}
\end{figure}

Accretional growth proportional to the instantaneous condensation
mass is a plausible but unproven assumption.
Myers (2000) used such a relation for the growth of condensations, 
but had to assume a very specific rate of growth of internal turbulence
as accretion proceeded.

Here, we look at an alternative scenario, also based on Bonnor-Ebert 
spheres, but assuming that the internal velocity dispersion remains
fixed during accretion. It is instructive to plot the radius versus mass
($R-M$) relation for the sequence of equilibrium states of increasing mass, 
but with fixed internal sound speed $\cs$ and external pressure $\Pe$. 
For small clouds, the $R-M$ relation is found analytically from the
Bonnor-Ebert theory (Bonnor 1956; Ebert 1957) to be
\beq
\label{rmrelation}
R = \left[ \frac{3 \, \cs^2}{4 \pi \Pe} \right]^{1/3} M^{1/3},
\eeq
However, this approximation also yields radii which are within 20\% of the
actual value for all $M/\Mu < 1$, where
$\Mu = \cs^4/(G^{3/2}\Pe^{1/2})$ is the natural unit of mass in the
Bonnor-Ebert problem.
Figure 3 shows the Bonnor-Ebert $R-M$ sequence as well as the corresponding 
relationship for a higher external pressure $3\Pe$.

In this simplified model, condensations start out somewhere on the
equilibrium $R-M$ sequence, increase their size $R$ according to the curve
as $M$ grows by accretion, then make a jump downward after an 
ambient pressure fluctuation, as shown schematically by the solid line
in Figure 3. After the downward jump, no accessible equilibrium state may
be present, and the condensation begins dynamical collapse.

If $R$ exceeds the Bondi (1952) radius
$R_{\rm B} \equiv G M/c_{\rm s,ext}$, where $c_{\rm s,ext}$ is the
sound speed in the external medium, then a stationary condensation 
undergoes geometric accretion
\beq
\label{geomacc}
\frac{dM}{dt} = 4 \pi R^2 \rho_{\rm ext} \, c_{\rm s,ext}. 
\eeq
Equations (\ref{rmrelation}) and (\ref{geomacc}) then lead to 
\beq
\label{bemdotandsoln}
\frac{dm}{dt} = \gamma_1 \, m^{2/3}  \: \: \Rightarrow \: \;
m(t) = \left[ m_0^{1/3} + \frac{\gamma_1}{3} \, t \right]^3,
\eeq
where $m = M/\Mu$ is a normalized mass, $m_0$ is the 
initial normalized mass, and $\gamma_1 = (36 \pi)^{1/3}\td^{-1} = 
4.84\,\td^{-1}$ is a growth rate.

Figure 4 shows the evolution of the probability density function $f(m)$
from an initial lognormal distribution (same as in Fig. 2) to a later 
distribution under the assumptions (1) all condensations grow for
a fixed time $t = \gamma_1^{-1}$, 
and (2) condensations grow for times that are distributed exponentially, 
with ``death'' rate 
$\delta = \gamma_1$. 
The first case leads to a profile that resembles a lognormal,
but is narrower than the initial distribution, and the second case
leads to a distinctly non-lognormal profile which may be described
as a broken power-law with two distinct regions; best-fitting slopes 
$\Gamma = d \log N/d \log m$ are equal to 
$-1.0$ in the intermediate mass range ($1 \leq m < 10$), $-2.4$
in the high mass range ($10 \leq m < 100$), and $-1.6$ for all
$m \geq 1$.
Interestingly, a broken power-law with a steeper slope at high
masses is reminiscent of some descriptions of the field-star IMF
(Kroupa \& Weidner 2003).

\section{Conclusions}

We believe that the
basic features of this model will be preserved in more complicated and 
realistic scenarios; that is, even a lognormal
distribution of condensation masses at birth coupled with subsequent
accretion growth (proportional to some power of mass) in which the 
accretion time obeys its own distribution (approximately exponential),
will result in a power-law tail.

Zinnecker (1982) has presented another model of accretion growth
in which a distribution of initial masses $m_0$ undergo Bondi
accretion ($\dot{m} \propto m^2$); the exponent greater than 1 means that
small differences in the $m_0$ values are magnified over time. After
a fixed accretion time for all objects, when $m \gg m_0$, $f(m)$ tends
to a power-law form with index $-2$.
For comparison, the mechanism of exponential distribution 
of lifetimes studied in this paper yields a power-law tail
even if the initial masses are all the same. The general combination 
of accretion growth (due to the Bondi process or
geometric accretion) and a distribution of accretion lifetimes
(where both processes are controlled by the dynamical time of the parent
system) may play a role in determining the IMF in detailed global 
simulations of star formation (e.g., Bate, Bonnell, \& Bromm 2003) in 
which stars are formed throughout the duration of the simulation.
Other ideas related to the {\it spatial} structure of turbulence may also
contribute to power-law fragment masses, e.g., thresholded regions in
density fields with lognormal probability density 
functions may have power-law clump
mass distributions (Pudritz 2002; Elmegreen 2002). However, we believe that
the {\it temporal} effect studied in this paper provides a compelling 
part of the explanation for the power-law tail in the IMF.

\section*{Acknowledgments}
SB was supported by a grant from the Natural Sciences and Engineering
Research Council of Canada (NSERC). CEJ acknowledges an 
NSERC Postdoctoral Fellowship.


\end{document}